\documentclass{article} 

\usepackage{PRIMEarxiv}
\usepackage{graphicx} 
\usepackage{epstopdf}

\usepackage{caption}
\usepackage{subcaption}
\usepackage[utf8]{inputenc} 
\usepackage[T1]{fontenc}    
\usepackage{hyperref}       
\usepackage{url}            
\usepackage{booktabs}       
\usepackage{amsfonts}       
\usepackage{nicefrac}       
\usepackage{microtype}      
\usepackage{lipsum}
\usepackage{amsmath}
\usepackage{multirow}
\usepackage{fancyhdr}       
\usepackage{graphicx}       
\graphicspath{{media/}} 

\usepackage{amsmath,amsfonts,bm}

\def\eqref#1{equation~\ref{#1}}

\def\1{\bm{1}}

\DeclareMathAlphabet{\mathsfit}{\encodingdefault}{\sfdefault}{m}{sl}
\SetMathAlphabet{\mathsfit}{bold}{\encodingdefault}{\sfdefault}{bx}{n}

\usepackage{enumitem}

\title{Meta-LegNet: A Transferable and Interpretable Framework for Surface Adsorption Prediction via Self-Defined Adsorption-Environment Learning}

\author{
  Li Yifan \\
  Department of Mechanical\\
  Engineering,National University\\
  of Singapore, 117575, SG\\
  \texttt{liyifan@nus.edu.sg} \\
  \And
    Arravind Subramanian
 \\
  Department of Chemical and Biomolecular\\
  Engineering,National University\\
  of Singapore, 117575, SG\\
  \texttt{arravind.subramanian07@u.nus.edu} \\
\And
   Liu Xiaoqing
 \\
  Department of Mechanical\\
  Engineering,National University\\
  of Singapore, 117575, SG\\
  \texttt{xqliu@nus.edu.sg} \\
\And
     Lyu Qiujie \\
  Department of Mechanical\\
  Engineering,National University\\
  of Singapore, 117575, SG\\
  \texttt{e1110106@u.nus.edu} \\
    \And
   Sergey Kozlov* \\
  Department of Chemical and Biomolecular\\
  Engineering,National University\\
  of Singapore, 117575, SG\\
  \texttt{sergey.kozlov@nus.edu.sg} \\
\And
   Shen Lei*\\
  Department of Mechanical\\
  Engineering,, National University\\
  of Singapore, 117575, SG\\
  \texttt{shelei@nus.edu.sg} \\
}

\begin{document}

\maketitle

\begin{abstract}
A central challenge in computational catalysis is the identification of low-energy and chemically plausible adsorption configurations, as these directly affect adsorption energies, reaction pathways, and catalytic performance. Existing
approaches generally rely on enumerating candidate adsorption sites followed by iterative refinement through density functional theory calculations or machine-learning-based relaxations. However, such workflows remain computationally
expensive and are difficult to scale to complex surfaces or multi-adsorbate systems. Here, we introduce Meta-LegNet, a graph learning framework that combines SE(3)-equivariant atom-level message passing with voxel-based multiscale aggregation and cross-domain meta-learning to learn transferable representations of local adsorption environments across diverse catalyst--adsorbate systems. Rather than following a conventional regression-only paradigm, Meta-LegNet encodes local chemical environments using invariant radial features and equivariant directional information, and further incorporates broader structural context through coordinate-frame voxel pooling, assignment-based upsampling, and gated feature fusion. The resulting local-global decomposition produces atom-resolved attribution maps, which are
processed to identify adsorption-relevant local environments in an interpretable manner. Based on the learned representations, we further construct an adsorption-environment database and develop a template-matching strategy to propose likely adsorption sites on previously unexplored surfaces without exhaustive site enumeration. Overall, our
results suggest that learning transferable adsorption environments provides an accurate, interpretable, and practical route for accelerating catalyst screening.
\end{abstract}

\section{Introduction}

In a typical computational workflow, an adsorbate is first placed on a catalyst surface, and then the resulting structure is relaxed using density functional theory (DFT) or machine-learning force fields (MLFFs) before any catalytic properties are evaluated.\cite{norskov2011dft,schutt2021mlff,mou2023bridging} Since these downstream quantities are highly sensitive to the final relaxed geometry, the construction of a reasonable adsorption structure is often the starting point for catalyst screening and mechanistic analysis.\cite{norskov2011dft,pettersson2022adsorption}

From the perspective of surface chemistry, this problem is closely connected to the concept of the active site. \cite{vogt2022active} In many catalytic systems, performance is not governed by the material as a whole, but rather by a limited number of local surface environments that control how reaction intermediates adsorb, stabilize, and transform. \cite{vogt2022active,mou2023bridging} Therefore, the key factor is not only the global identity of the catalyst, but also the local geometric and chemical structure surrounding the adsorption site. \cite{vogt2022active,usuga2024local} For many systems, the dominant information required to describe adsorption behavior is already contained within a relatively small group of nearby atoms, suggesting that transferable local representations may provide an effective way to characterize adsorption environments across diverse surfaces. \cite{usuga2024local,li2023local}

However, determining which local environment should be considered remains a major challenge. \cite{mou2023bridging,wang2022mladsorption} Many existing atomistic models define local neighborhoods using fixed distance cutoffs, nearest-neighbor rules, or other hand-crafted criteria. \cite{schutt2021mlff,wang2022mladsorption} Although such strategies can be effective for limited classes of materials, their reliability often decreases when surface composition, morphology, or adsorbate identity changes. \cite{mou2023bridging,wang2022mladsorption} This difficulty becomes even more pronounced in practical adsorption searches, where relevant local environments must be identified before they can be evaluated. Even for simple surfaces, one typically needs to enumerate atop, bridge, hollow, and multiple orientation-dependent variants.\cite{lan2023adsorbml,hoffmann2022automated} When multiple adsorbates are present, the number of possible configurations increases combinatorially; for \(m\) candidate sites and \(n\) adsorbates, the search space scales as \(O(m^n)\).

\begin{figure*}[t]
\centering
\includegraphics[width=1\textwidth]{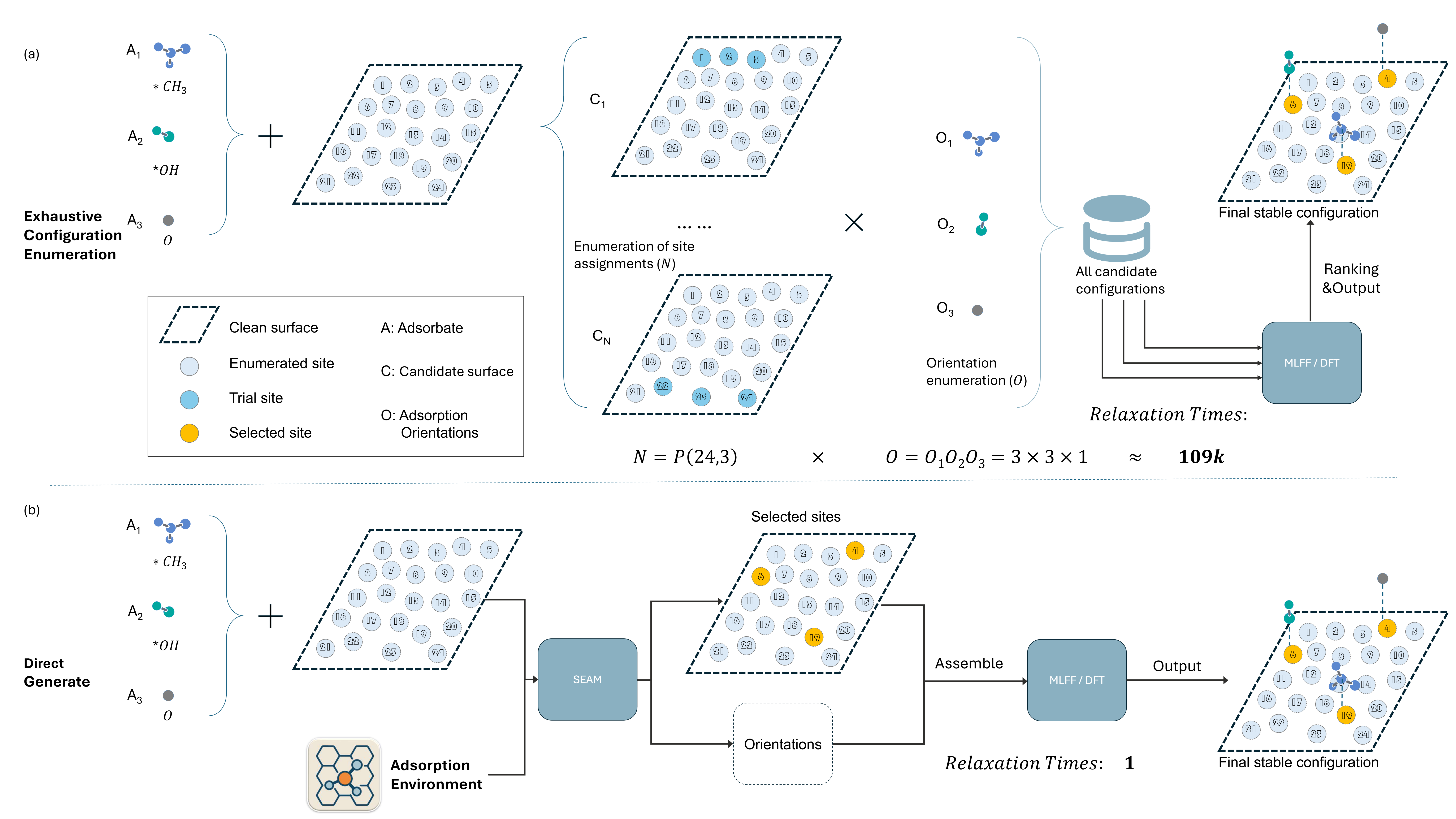}
\caption{Comparison between the conventional sequential adsorption workflow and the proposed direct adsorption workflow based on Site Extraction via Adsorption-environment Matching (SEAM). (a) In a conventional DFT-based workflow, adsorbates \(A_1, A_2, \ldots, A_n\) are added to enumerated surface sites and relaxed iteratively. Each additional adsorbate requires a new adsorption and relaxation cycle. (b) In SEAM, the clean slab and all adsorbates are provided together, and adsorption-environment matching directly proposes feasible sites and poses in a single pass.}
\label{fig:Schematic}
\end{figure*}

To address this issue, a common strategy in high-throughput catalysis studies is to enumerate a large number of candidate adsorption structures and relax each of them using DFT.\cite{norskov2011dft,chanussot2021open,lan2023adsorbml} While this approach is conceptually straightforward, its computational cost remains extremely high. Each structural relaxation requires repeated electronic-structure calculations together with iterative geometry optimization, and a single relaxation may involve many to hundreds of single-point evaluations depending on the system and convergence behavior.\cite{chanussot2021open,mou2023bridging} The scale of the OC20 dataset provides a representative example: its training split for the S2EF task is sampled from 640,081 relaxations, and the full benchmark contains 1,281,040 DFT relaxations and approximately 264,890,000 single-point evaluations.\cite{chanussot2021open}

Machine-learning models can significantly reduce the cost of individual energy and force evaluations, but in most cases they do not fundamentally alter this overall workflow. \cite{chanussot2021open,schutt2021mlff,wang2022mladsorption} In structure-to-energy-and-force settings, the model still needs to be called repeatedly during relaxation, meaning that the total cost and accumulated error remain coupled to a sequential optimization process. \cite{chanussot2021open,schutt2021mlff} In the initial-structure-to-relaxed-energy (IS2RE) setting, the relaxation trajectory is bypassed, but the prediction target remains difficult because it depends on a final adsorption geometry that is not known in advance.\cite{chanussot2021open} More importantly, both settings generally assume that a reasonable candidate adsorption structure has already been generated.\cite{chanussot2021open,lan2023adsorbml} As a result, existing surrogate models mainly accelerate the scoring or relaxation of candidate structures after they have been proposed, rather than addressing the more fundamental question of how to identify transferable adsorption environments and directly construct plausible adsorption configurations.\cite{mou2023bridging,lan2023adsorbml}

In this work, we introduce Meta-LegNet, a graph learning framework with SE(3)-equivariant atom-level message passing for learning transferable representations of adsorption environments across diverse catalyst and adsorbate systems. Rather than treating adsorption solely as a direct structure-to-property prediction problem, Meta-LegNet is designed to learn reusable representations of local adsorption environments through invariant radial features, equivariant directional message passing, voxel-based multiscale aggregation, gated feature fusion, and cross-domain meta-learning. Based on these learned representations, we further organize an adsorption-environment database and develop a Site Extraction
via Adsorption-environment Matching (SEAM) procedure, which proposes plausible adsorption sites on previously unseen surfaces and constructs candidate adsorption configurations without exhaustive site enumeration.

\begin{figure*}[t]
\centering
\includegraphics[width=1\textwidth]{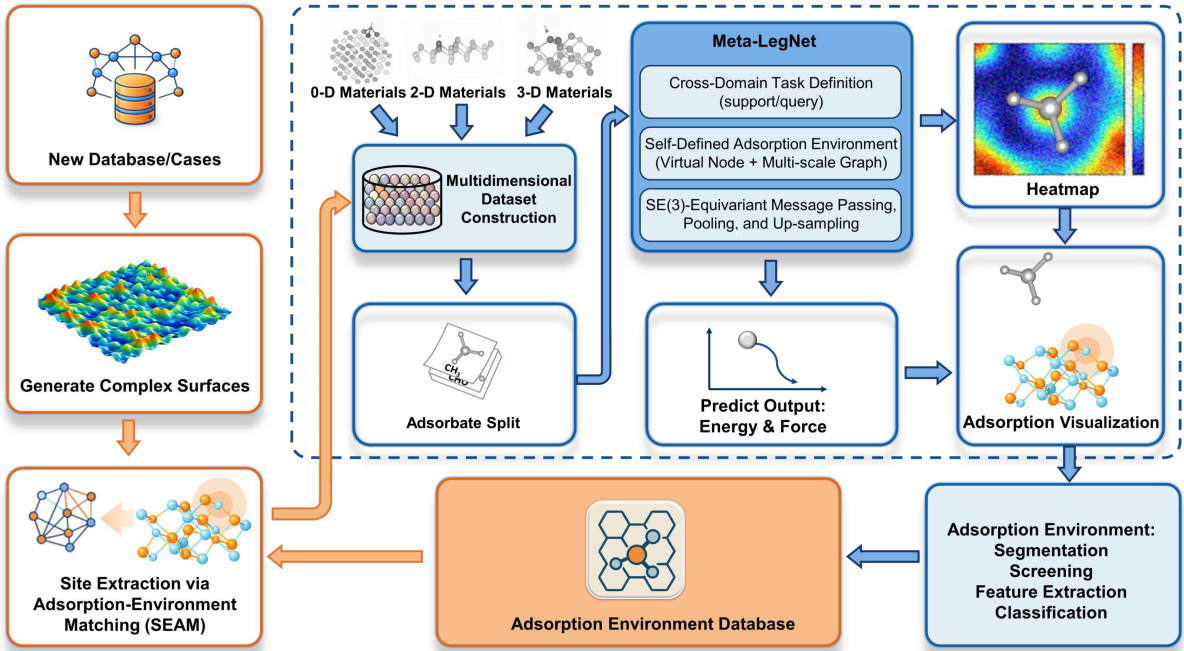}
\caption{Overview of Meta-LegNet. The framework integrates multidimensional
adsorption datasets spanning 0D, 2D, and 3D materials; defines cross-domain meta-learning tasks using support and query splits; learns local adsorption representations through SE(3)-equivariant atom-level message passing,
voxel-based multiscale aggregation, assignment-based upsampling, and gated fusion; and outputs adsorption energies, optional force predictions, and atom-wise attribution maps. The attribution maps are further processed, and the resulting environments are organized into a reusable adsorption-environment database for SEAM-based adsorption-site proposal.}
\label{fig:main}
\end{figure*}

\section{Materials and Methods}

\subsection{Framework Overview}

Meta-LegNet is designed with two goals in mind: (i) to learn transferable local adsorption representations from heterogeneous adsorption data, and (ii) to reuse these representations for prediction and site localization on new
catalyst--adsorbate systems. The framework proceeds in three stages. First, each adsorption structure is represented as an atomic graph whose edges are constructed from interatomic distances. Atom-level scalar and vector features are updated by SE(3)-equivariant message-passing blocks using invariant radial embeddings and equivariant directional information. Broader structural context is then incorporated through voxel-based multiscale pooling, assignment-based upsampling, and gated feature fusion. Second, the resulting multiscale representation is optimized in a cross-domain meta-learning setting so that the model can adapt to new tasks with limited support data. Third, the learned adsorption-environment representations are extracted, examined, and organized into a database that supports direct adsorption-site proposal through SEAM. This design shifts the focus from exhaustive candidate enumeration to learning reusable adsorption-environment representations that can transfer across systems.

\subsection{Datasets}

We constructed a multidimensional adsorption benchmark spanning 0D, 2D, and 3D catalytic systems. The benchmark was designed to cover diverse adsorption motifs, surface chemistries, and structural dimensionalities, thereby providing a demanding test of model transferability.

\paragraph{Zero-Dimensional Nanocluster Adsorption Database.}
This component contains two subsets. The first includes 1,804 distinct adsorption configurations involving 38 adsorbates. The second provides adsorption trajectories for five selected adsorbates, yielding 65,806 individual structures.

\paragraph{Two-Dimensional Materials Database.}
The 2D database contains 2,803 DFT-calculated hydrogen adsorption energies on a wide range of two-dimensional surfaces \cite{yang2020high}. All calculations are based on surface structures from the 2DMatPedia database, which contains more than 10,000 unique 2D materials \cite{zhou20192dmatpedia}.

\paragraph{Three-Dimensional Materials Dataset.}
The 3D collection includes a binary alloy dataset with 53,840 samples\cite{mamun2019high}, a subset of 159,843 OC20-derived samples covering 16 common adsorbates\cite{chanussot2021open}, and 50,000 oxide-surface adsorption configurations sampled from OC22\cite{tran2023open}.

Together, these datasets cover nanoclusters, layered materials, alloys, and oxides, and they expose the model to substantial variation in composition, structure, and adsorption behavior. Splits were made at the trajectory level, not at the frame level, so that no relaxation trajectory contributed frames to more than one of train, validation and test.

\begin{figure*}[t]
\centering
\includegraphics[width=0.8\textwidth]{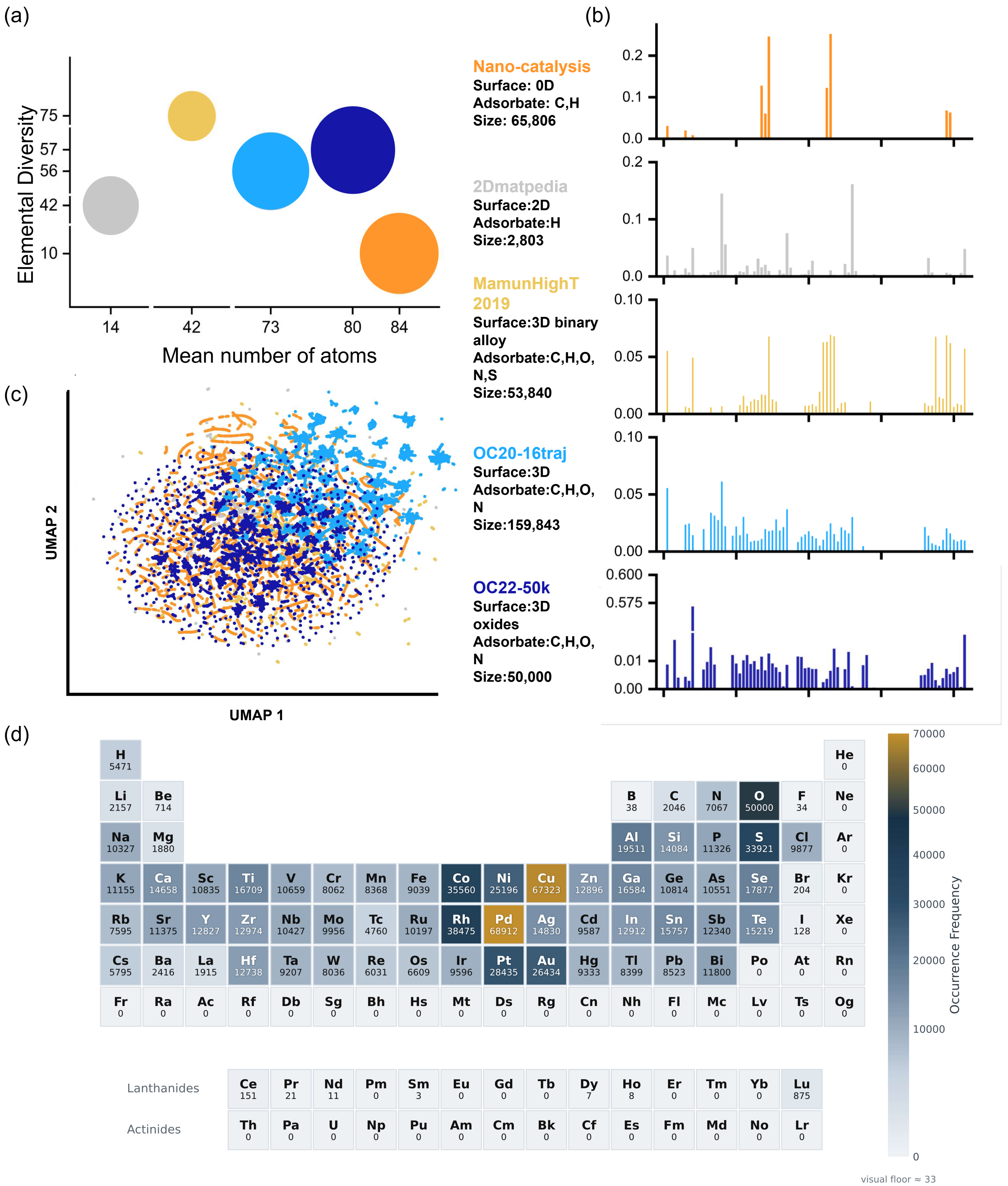}
\caption{Statistical overview of the adsorption benchmark.
(a) Elemental diversity and mean number of atoms for each dataset; bubble size scales with dataset size.
(b) Elemental occurrence probabilities of surface species across datasets.
(c) UMAP projection of surface-adsorbate pairs from joint SOAP and RDKit embeddings, showing complementary but partially overlapping coverage of structural and chemical space across 0D, 2D and 3D surface systems.
(d) Periodic-table representation of elemental frequencies across the benchmark}
\label{fig:DATASET}
\end{figure*}

\subsection{Graph Representation and Self-Defined Adsorption-Environment Learning}

Conventional local-environment modeling in atomistic learning often relies on hand-crafted neighborhood definitions, such as fixed distance cutoffs, nearest-neighbor sorting, or Voronoi-based rules. Although effective in many settings, these choices can introduce sensitivity to local density variations and discretization details, which may reduce the stability of the learned representation.

In Meta-LegNet, each adsorption structure is represented as an equivariant atomic graph \(G=(V,E)\), where \(V\) and \(E\) denote atoms and local interactions, respectively. For each atom \(i\), we define scalar features \(h_i \in \mathbb{R}^{F}\), vector features \(\mathbf{v}_i \in \mathbb{R}^{3 \times F}\), and Cartesian coordinates \(\mathbf{r}_i \in \mathbb{R}^{3}\). The scalar features are initialized from atomic numbers using a learnable embedding, while the vector features are initialized as zeros. The edge set is constructed by a radius graph, and interatomic distances are encoded through radial basis functions together with normalized relative directions.

To improve the robustness of local-environment encoding, we introduce a multiscale representation based on equivariant voxel pooling rather than repeated heuristic redefinition of neighborhoods. After atom-level message passing, atoms within the same spatial voxel are aggregated into coarse nodes. Their scalar, vector, and positional information is pooled in an equivariant manner, and the resulting coarse graph is processed again to capture broader structural context. Repeating this procedure yields a hierarchical description of the adsorption environment in which fine-scale atomic interactions and coarser structural context are encoded jointly.

To preserve local detail, the pooled representations are projected back to the atom level through equivariant upsampling, and features from different resolutions are combined by gated fusion. This coarse-to-fine interaction allows the model to refine local descriptors using broader structural context while maintaining atom-level resolution for downstream prediction. In addition, lightweight virtual-node embeddings are introduced to facilitate global information exchange across the graph.

Overall, Meta-LegNet learns the adsorption environment through the coordinated use of atom-level equivariant message passing, hierarchical voxel-based pooling, cross-scale feature fusion, and graph-level information exchange. This design allows the model to capture stable and expressive adsorption patterns across spatial scales.

\subsection{Meta-LegNet Architecture}

The architecture has two coupled components: (i) SE(3)-equivariant atom-level message passing, which updates scalar and vector features using invariant radial embeddings and equivariant directional information, and (ii) voxel-based multiscale
graph coarsening, which aggregates local representations into progressively coarser structural contexts. The voxel coarsening is translation-consistent in the simulation coordinate frame and its pooling operator is equivariant conditional on a fixed coarse assignment. We therefore distinguish the strictly equivariant neural update blocks from the coordinate-frame-dependent voxel assignment used for efficient multiscale aggregation.

\begin{figure*}[t]
\centering
\includegraphics[width=1\textwidth]{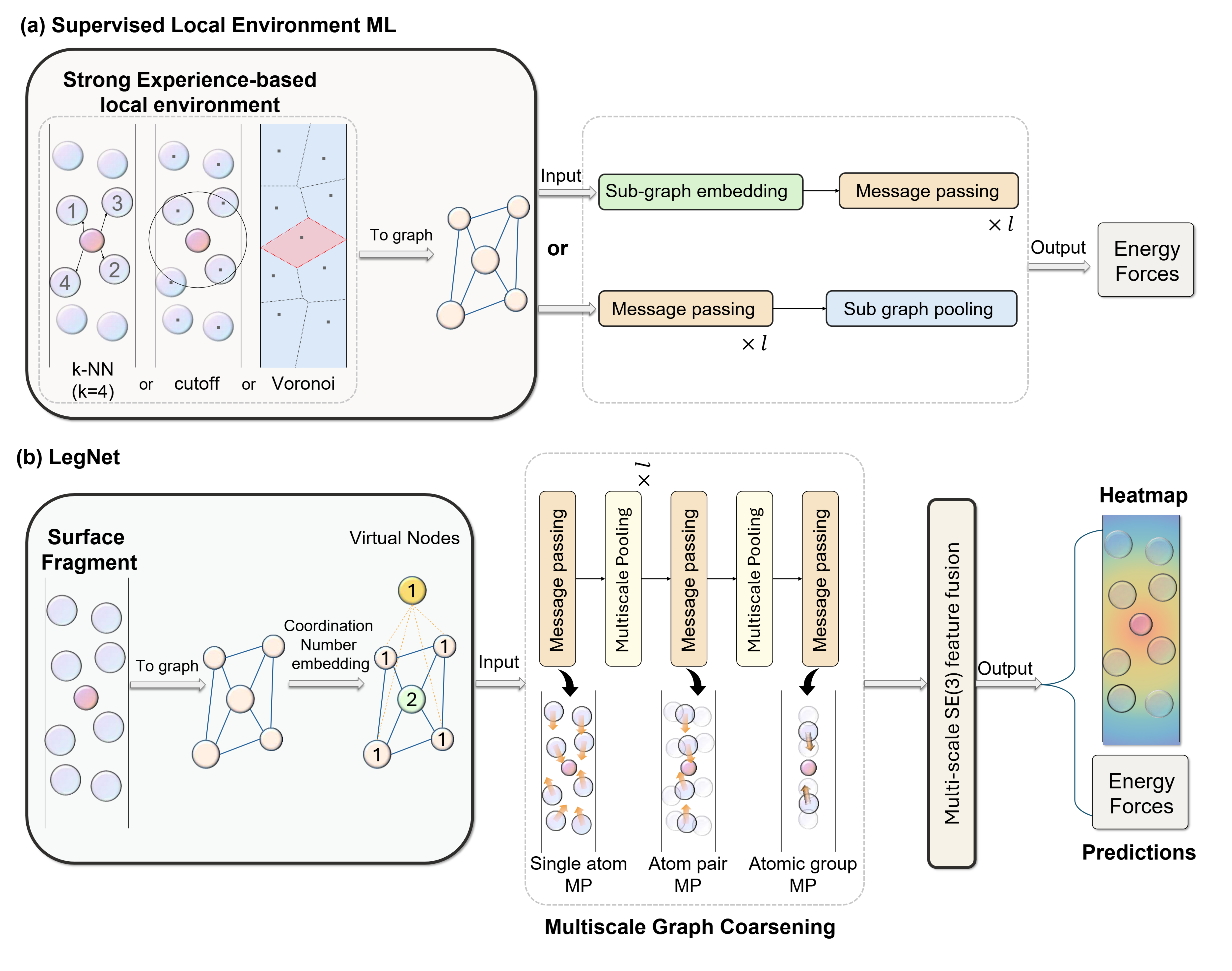}
\caption{Architecture of Meta-LegNet. Starting from atomistic graphs, the model combines self-defined adsorption-environment learning, multiscale equivariant message passing, hierarchical graph coarsening, and cross-domain meta-learning to produce transferable representations across catalytic systems.}
\label{fig:workflow}
\end{figure*}

\begin{figure*}[t]
\centering
\includegraphics[width=1.0\textwidth]{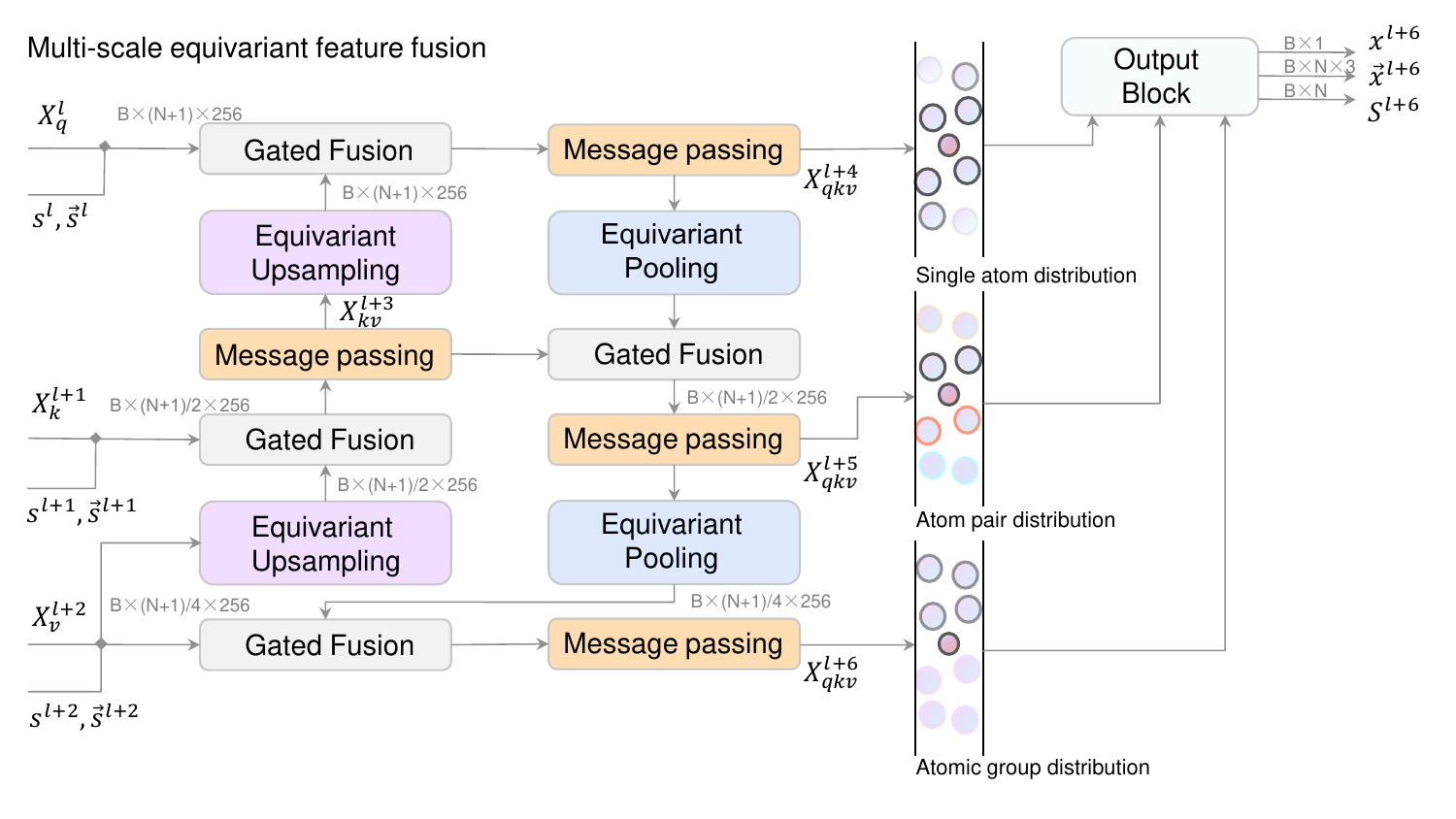}
\caption{Multiscale equivariant feature fusion in Meta-LegNet. Coarse-grained graph features provide broader structural context, while upsampled fine-grained features preserve local atomic detail. A gated fusion mechanism combines both scales before the final prediction.}
\label{fig:model}
\end{figure*}

\subsubsection{Equivariant Message Passing}

Let \(h_i^{(l)} \in \mathbb{R}^{F}\) and
\(\mathbf{v}_i^{(l)} \in \mathbb{R}^{3 \times F}\) denote the scalar and vector
features of node \(i\) at layer \(l\). For an edge \((i,j)\), we define
\[d_{ij}=\|\mathbf{r}_j-\mathbf{r}_i\|_2,
\qquad\hat{\mathbf{r}}_{ij}=\frac{\mathbf{r}_j-\mathbf{r}_i}{\|\mathbf{r}_j-\mathbf{r}_i\|_2}.\]
The distance \(d_{ij}\) is encoded by a radial basis embedding \(\phi(d_{ij})\).

In our implementation, scalar coefficients are generated from invariant quantities, including node scalar features and radial distance embeddings. The vector message is formed as a linear combination of equivariant vector bases:
\begin{equation}
\mathbf{m}_{ij}^{(l)}=\alpha_{ij}^{(l)} \mathbf{v}_j^{(l)}+\beta_{ij}^{(l)} \hat{\mathbf{r}}_{ij},
\end{equation}
where \(\alpha_{ij}^{(l)}\) and \(\beta_{ij}^{(l)}\) are scalar coefficients produced from invariant inputs. The scalar message is written as
\begin{equation}
m_{ij}^{(l)}=\gamma_{ij}^{(l)},
\end{equation}
where \(\gamma_{ij}^{(l)}\) is also produced from invariant scalar inputs.

The node updates are then
\begin{align}
h_i^{(l+1)}&=h_i^{(l)}+\sum_{j\in\mathcal{N}(i)}m_{ij}^{(l)},\\
\mathbf{v}_i^{(l+1)}&=\mathbf{v}_i^{(l)}+\sum_{j\in\mathcal{N}(i)}\mathbf{m}_{ij}^{(l)}.
\end{align}

Under a rigid transformation
\[\mathbf{r}_i \mapsto R\mathbf{r}_i+\mathbf{t},\qquad\mathbf{v}_i \mapsto R\mathbf{v}_i,\]
the distance \(d_{ij}\) and the radial embedding \(\phi(d_{ij})\) remain invariant, while \(\hat{\mathbf{r}}_{ij}\mapsto R\hat{\mathbf{r}}_{ij}\). Therefore \(\mathbf{m}_{ij}^{(l)}\mapsto R\mathbf{m}_{ij}^{(l)}\), while \(m_{ij}^{(l)}\) remains invariant. Summation over neighbors preserves these transformation rules. Thus, the message-passing update is SE(3)-equivariant up to the usual numerical tie-breaking in radius-based neighbor construction.

\subsubsection{Voxel-Based Pooling and Conditional Equivariance}

The multiscale coarsening module assigns atoms to voxels in the simulation coordinate frame. For a graph \(b\), let \(\mathbf{o}_b\) denote a graph-level origin, chosen as either the coordinate-wise minimum or the coordinate mean. With voxel size \(s\), the voxel index of atom \(i\) is
\begin{equation}
\mathbf{g}_i=\left\lfloor\frac{\mathbf{r}_i-\mathbf{o}_{b_i}}{s}\right\rfloor .
\end{equation}
Atoms with the same voxel key are assigned to the same coarse node. Let \(q(i)\) denote this assignment and
\[\mathcal{C}_p=\{i\in V \mid q(i)=p\}\]
denote the set of atoms assigned to coarse node \(p\).

For a fixed assignment \(q\), pooled scalar, vector, and coordinate features are computed by mean aggregation:
\begin{align}
h_p&=\frac{1}{|\mathcal{C}_p|}\sum_{i\in\mathcal{C}_p}h_i,\\
\mathbf{v}_p&=\frac{1}{|\mathcal{C}_p|}\sum_{i\in\mathcal{C}_p}\mathbf{v}_i,\\
\mathbf{r}_p&=\frac{1}{|\mathcal{C}_p|}\sum_{i\in\mathcal{C}_p}\mathbf{r}_i .
\end{align}

If the assignment \(q\) is unchanged under a rigid transformation, then the pooling operator satisfies
\begin{align}
h_p' &= h_p,\\\mathbf{v}_p' &= R\mathbf{v}_p,\\\mathbf{r}_p' &= R\mathbf{r}_p+\mathbf{t}.
\end{align}
Thus, the aggregation operator itself is SE(3)-equivariant conditional on a fixed coarse assignment.

However, the voxel assignment used here is defined on a fixed Cartesian grid in the simulation coordinate frame. It is translation-consistent because translating all coordinates also translates the graph-level origin, leaving \(\mathbf{r}_i-\mathbf{o}_{b_i}\) unchanged. It is not strictly invariant to arbitrary rotations, because rotating the structure can change the Cartesian voxel indices.
Accordingly, the voxel module should be interpreted as an efficient coordinate-frame multiscale aggregation scheme rather than a strictly rotation-equivariant coarsening operator.

\subsubsection{Upsampling}

Upsampling maps coarse features back to atom-level nodes using the stored assignment \(q\). For \(i\in\mathcal{C}_p\), we define
\begin{align}
h_i^{\mathrm{up}} &= h_p,\\\mathbf{v}_i^{\mathrm{up}} &= \mathbf{v}_p .
\end{align}
The original fine-scale coordinates are retained for subsequent atom-level operations.

For a fixed assignment \(q\), upsampling preserves the transformation properties of the coarse features:
\[h_i^{\mathrm{up}\,\prime}=h_i^{\mathrm{up}},\qquad\mathbf{v}_i^{\mathrm{up}\,\prime}=R\mathbf{v}_i^{\mathrm{up}}.\]
Therefore, upsampling is equivariant conditional on the same assignment used during pooling. In the current voxel implementation, this condition is exact for translations but not guaranteed for arbitrary rotations because voxel membership may change after rotation.

\subsubsection{Gated Multiscale Fusion}

Coarse and fine scalar features are fused using a gate computed from scalar representations:
\begin{equation}
g_i=\sigma\left(W_g [h_i^{\mathrm{up}};h_i^{\mathrm{fine}}]\right).
\end{equation}
The fused scalar representation is
\begin{equation}
\tilde{h}_i=g_i\odot h_i^{\mathrm{up}}+(1-g_i)\odot h_i^{\mathrm{fine}}.
\end{equation}

The same invariant scalar gate is used to combine vector features:
\begin{equation}
\tilde{\mathbf{v}}_i=g_i\odot \mathbf{v}_i^{\mathrm{up}}+(1-g_i)\odot \mathbf{v}_i^{\mathrm{fine}}.
\end{equation}
Because \(g_i\) is computed from scalar features and is therefore invariant, and because both vector inputs transform equivariantly, the fused vector feature also transforms equivariantly for a fixed voxel assignment.

\subsubsection{Output Block}

The final atom-level scalar representation \(h_i^{(L)}\) is used to predict a learned local contribution and a learned scalar gate:
\begin{align}
e_i^{\mathrm{raw}}&=f_{\mathrm{loc}}(h_i^{(L)}),\\
g_i&=\sigma\left(f_{\mathrm{gate}}(h_i^{(L)})\right).
\end{align}
The effective node-level contribution is
\begin{equation}
c_i=e_i^{\mathrm{raw}}g_i .
\end{equation}

A graph-level correction is computed from the mean-pooled scalar representation:
\begin{equation}
e_{\mathrm{glob}}=f_{\mathrm{glob}}\left(\frac{1}{|V|}\sum_{i\in V}h_i^{(L)}\right).
\end{equation}
The final adsorption-energy prediction is
\begin{equation}
\hat{E}_{\mathrm{ads}}=\sum_{i\in V}c_i+e_{\mathrm{glob}}.
\end{equation}

Because this readout uses scalar node features and permutation-invariant aggregation,
it is invariant to atom ordering. It is also invariant to rigid motions provided that
the preceding representation is invariant, which holds exactly for the equivariant
message-passing blocks and conditionally for the voxel-based multiscale blocks when
the coarse assignment remains unchanged.

For force prediction, the implementation supports two modes. In the default mode,
forces are predicted from the final equivariant vector features using a linear
vector readout:
\begin{equation}
\hat{\mathbf{F}}_i
=
W_F \tilde{\mathbf{v}}_i .
\end{equation}
Since \(W_F\) acts only on feature channels and not on spatial coordinates, this
readout is rotation-equivariant when \(\tilde{\mathbf{v}}_i\) is equivariant.
However, this independent force head does not by itself guarantee a conservative
force field.

When conservative forces are required, the model can instead compute forces from the
energy gradient:
\begin{equation}
\hat{\mathbf{F}}_i
=
-
\frac{\partial \hat{E}_{\mathrm{ads}}}{\partial \mathbf{r}_i}.
\end{equation}
This mode enforces energy-consistent forces within regions where the graph
connectivity and voxel assignment remain fixed.

\subsection{Meta-Learning Framework}

Each adsorption prediction scenario is treated as a meta-task. A task corresponds to a specific catalyst--adsorbate setting with its own support and query sets. This formulation is motivated by the fact that adsorption behavior can vary substantially across materials and adsorbates, while still sharing transferable local structural patterns.

\begin{figure*}[t]
\centering
\includegraphics[width=1.0\textwidth]{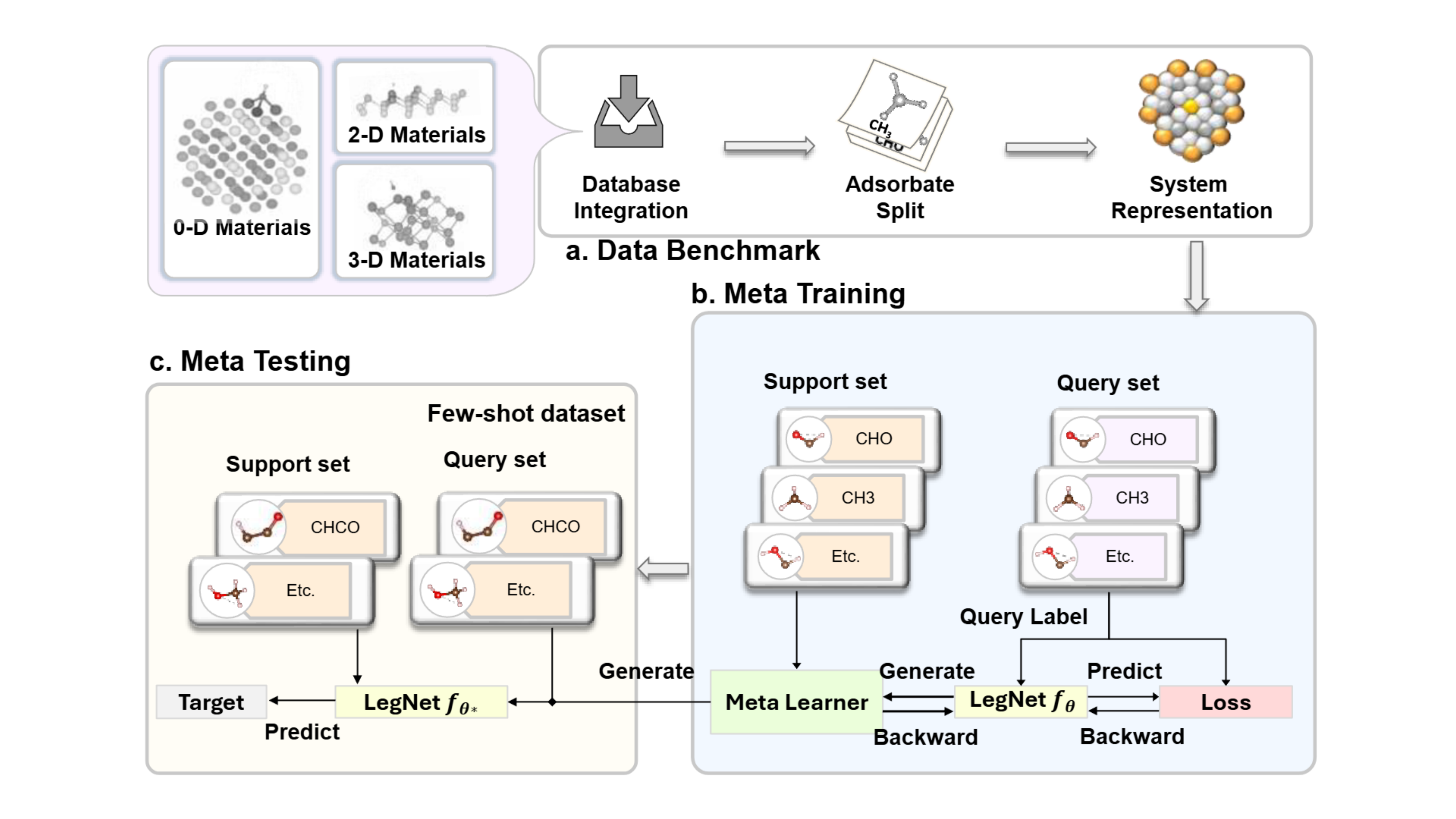}
\caption{Cross-domain meta-learning in Meta-LegNet. A reorganized benchmark spanning multidimensional adsorption systems is split into source and target domains. During meta-training, transferable knowledge is learned through bi-level optimization over source-domain tasks. During meta-testing, the learned initialization is adapted to new adsorption tasks in the target domain with limited support data.}
\label{fig:META}
\end{figure*}

We adopt the Model-Agnostic Meta-Learning (MAML) algorithm \cite{finn2017model}. Let \(\mathcal{D}_{\text{source}}\) denote the source domain, which contains diverse labeled adsorption tasks, and let \(\mathcal{D}_{\text{target}}\) denote the target domain, which contains previously unseen scenarios with limited labels. In each meta-training episode, we sample \(m\) tasks \(\{\mathcal{T}_i\}_{i=1}^{m}\) from \(\mathcal{D}_{\text{source}}\). Each task is split into a support set \(\mathcal{D}_{i}^{\text{support}}\) and a query set \(\mathcal{D}_{i}^{\text{query}}\).

For task \(\mathcal{T}_i\), the inner update is
\begin{equation}
\theta_i' = \theta - \alpha \nabla_{\theta} \mathcal{L}_{\text{support}}^{(i)}(f_{\theta}),
\end{equation}
where \(\theta\) is the shared initialization, \(\alpha\) is the inner-loop learning rate, and \(\mathcal{L}_{\text{support}}^{(i)}\) is the loss on \(\mathcal{D}_{i}^{\text{support}}\). The outer update optimizes the initialization across the sampled tasks:
\begin{equation}
\theta \leftarrow \theta - \beta \nabla_{\theta} \sum_{i=1}^{m} \mathcal{L}_{\text{query}}^{(i)}\!\left(f_{\theta_i'}\right),
\end{equation}
where \(\beta\) is the outer-loop learning rate.

At meta-test time, the learned initialization is adapted to tasks sampled from \(\mathcal{D}_{\text{target}}\). With only a few support examples, the model can be fine-tuned and evaluated on the corresponding query set. This framework reduces the dependence on large task-specific datasets by transferring common structural and chemical patterns across adsorption environments.

\subsection{Adsorption-Guided Representation Learning}

Adsorption energy is defined as
\[E_{\mathrm{ads}}=E_{\mathrm{surface+adsorbate}}-E_{\mathrm{surface}}-E_{\mathrm{adsorbate}}\],
where a more negative value indicating stronger stabilization of the adsorbate.

Because adsorption is primarily governed by the local chemical and geometric environment around the binding site, we incorporate adsorption information through a learned local-contribution readout. After equivariant atom-level message passing, voxel-based multiscale aggregation, upsampling, and gated fusion, each atom \(i\) is associated with a scalar representation \(h_i\). The model predicts a raw local contribution and an adsorption gate:
\begin{align}
e_i^{\mathrm{raw}}&=f_{\mathrm{loc}}(h_i),\\
g_i&=\sigma\left(f_{\mathrm{gate}}(h_i)\right).
\end{align}
The effective node-level contribution is
\begin{equation}
c_i=e_i^{\mathrm{raw}}g_i .
\end{equation}

The final adsorption-energy prediction is
\begin{equation}
\hat{E}_{\mathrm{ads}}=\sum_{i\in V}c_i+f_{\mathrm{glob}}\left(\frac{1}{|V|}\sum_{i\in V}h_i\right).
\end{equation}
The first term captures learned local-environment contributions, while the second term models residual collective effects through a graph-level correction.

The quantities \(c_i\) are used as model-derived attribution scores for identifying atoms or local regions that strongly influence the prediction. They should not be interpreted as a unique physical decomposition of the adsorption energy, since such a decomposition is not uniquely defined.

\section{Results}

\subsection{Experimental Setup and Evaluation Metrics}

We evaluated Meta-LegNet on all five datasets using the adsorption-prediction settings considered in this work: structure-to-energy (S2E), structure-to-force (S2F), and initial-structure-to-relaxed-energy (IS2RE). The S2E and S2F tasks assess the model's ability to predict energies and forces for input structures, while IS2RE predicts the relaxed adsorption energy from an initial, unrelaxed configuration.

When both types of supervision are available, the model uses a shared backbone with task-specific heads so that force information can improve representation learning for energy prediction. Unless otherwise stated, performance is reported as mean absolute error (MAE), and lower values indicate better performance.

\subsection{Effect of Multi-Scale Pooling and Upsampling}

We next evaluate the role of multi-scale pooling and upsampling in adsorption- environment learning. Voxel-based pooling groups atom-level representations into coarser structural units, allowing the model to incorporate surface context beyond
the immediate bonding environment. This broader context is important because adsorption stability is often affected by the surrounding coordination structure, local morphology, and neighboring surface atoms rather than by the binding atom
alone.

Upsampling then transfers the coarse representations back to the original atom-level graph. This coarse-to-fine pathway refines local atomic descriptors with information from larger surface regions, enabling the model to combine short-range chemical interactions with broader structural context. The resulting representation therefore captures adsorption environments at multiple resolutions, from individual atoms to local clusters and coarse surface regions.

The multiscale module also has conditional equivariance. For a fixed voxel assignment, the pooling, assignment-based upsampling, and gated fusion operations preserve the transformation behavior of scalar and vector features: scalar features remain invariant, while vector features rotate consistently with the input coordinates. In the present implementation, the voxel assignment is defined in the simulation coordinate frame and therefore provides an efficient multiscale organization of the atomic graph.  This formulation provides a practical balance between geometric consistency, computational efficiency, and the ability to capture broader adsorption environments.

Ablation results in Supplementary Table~\ref{tab:ablation} show that removing this multi-scale pathway consistently increases the MAE. These results demonstrate that pooling and upsampling are not merely architectural additions, but provide a substantial contribution by expanding the effective receptive field, improving contextual representation, and strengthening atom-level adsorption attribution.

\subsection{Benchmark Performance}

Table~\ref{tab:extracted_results} summarizes the main benchmark results.On the OC20-16traj split, Meta-LegNet achieved an S2E MAE of 0.1895 eV, compared with 0.2326 eV for GemNet-dT and 0.2588 eV for DimeNet++. For S2F on the same split, Meta-LegNet achieved 0.0201 eV/Å. These values should be interpreted with uncertainty estimates and under the exact split definition reported in Methods

Similar gains are observed on the 0D trajectory dataset, where Meta-LegNet achieves \(0.10\) eV for S2E and \(0.0095\) eV/\AA\ for S2F. The advantage is especially clear on IS2RE, where the model obtains \(0.105\), \(0.026\), and \(0.097\) eV on the 0D, 2D, and 3D datasets, respectively. These results suggest that the learned adsorption-environment representation is useful both for local force-sensitive prediction and for extrapolating relaxed adsorption energies from unrelaxed inputs.

\begin{table}[htbp]
\centering
\caption{MAE comparison between Meta-LegNet and baseline models on representative adsorption benchmarks. Lower values indicate better performance.}
\label{tab:extracted_results}
\small
\setlength{\tabcolsep}{6pt}
\begin{tabular}{llccccccc}
\hline
Task & Split & Meta-LegNet & GemNet-dT & DimeNet++  & PAINN& ComENet & SchNet & CGCNN \\
\hline
S2E (eV) & OC20-16traj   & \textbf{0.1895}  & 0.2326  & 0.2588  & 0.2795  & 0.2719  & 0.4167  & 0.4974\\
         & OC22-50k       & \textbf{0.5012}  & 0.5412  & 0.5827  & 0.6325  & 0.6089  & 1.1048  & 1.3396 \\
         & Nano-catalysis  & \textbf{0.1013}  & 0.1271  & 0.1353 & 0.1284 & 0.1121 & 0.2015 & 0.2117 \\
\hline
S2F (eV/\AA) & OC20-16traj  & \textbf{0.0201} & 0.0213  & 0.0237  & 0.0256  & 0.0249 & 0.0290  & 0.0364 \\
             & OC22-50k      & \textbf{0.0365}  & 0.0387  & 0.0430  & 0.0464  & 0.0453  & 0.0527  & 0.0662 \\
             & Nano-catalysis & \textbf{0.0095} & 0.0105 & 0.0128 & 0.0143 & 0.0118 & 0.0132 & 0.0140 \\
\hline
IS2RE (eV) & OC20-16traj  & \textbf{0.3445} & 0.3878  & 0.4325  & 0.4698  & 0.5010 & 0.9642  & 1.2145 \\
           & OC22-50k      & \textbf{1.1874}  &1.4524  & 1.6266  & 1.5794  & 1.8506  & 3.1964  & 3.6409 \\
           & Nano-catalysis         & \textbf{0.1052} & 0.1323 & 0.1281 & 0.1427 & 0.1304 & 0.1335 & 0.1423 \\
           & 2Dmatpedia        & \textbf{0.0261}  & 0.0334  & 0.0328  & 0.0335  & 0.0425  & 0.0576 & 0.0422 \\
           & MamunHighT-2019         & \textbf{0.0968}  & 0.1254 & 0.1355 & 0.1664  & 0.1434  & 0.2683  & 0.2751 \\
\hline
\end{tabular}
\end{table}

Overall, the benchmark results support two conclusions. First, Meta-LegNet improves predictive accuracy across structurally diverse datasets rather than on a single benchmark only. Second, the gains are consistent across both structure-conditioned tasks (S2E/S2F) and relaxed-energy prediction (IS2RE), indicating that the learned local representation transfers across related adsorption objectives.

\subsection{Few-Shot Adaptation}

To test low-data generalization, we conducted \(1\)-, \(5\)-, \(10\)-, and \(20\)-shot experiments on three representative adsorption tasks of different dimensionalities:
\begin{enumerate}[label=(\roman*)]
    \item a 0D trimetallic cluster adsorbing CH\(_2\)CO for the S2F task;
    \item a 2D surface adsorbing CO\(_2\) for the IS2RE task; and
    \item a 3D system adsorbing CHOHCH\(_2\)OH for the S2F task.
\end{enumerate}

We compared Meta-LegNet with MAML-trained versions of CGCNN, SchNet, and ComENet. The data here is a small amount of additionally calculated data used to validate the meta-framework and therefore is not included in the dataset section. As shown in Table~\ref{tab:few_shot_results_three}, Meta-LegNet consistently outperforms all baselines across all shot regimes. The largest margin appears on the 3D CHOHCH\(_2\)OH task, where the \(1\)-shot MAE is \(0.33\) eV for Meta-LegNet, compared with \(0.69\), \(0.63\), and \(0.51\) eV for Meta-CGCNN, Meta-SchNet, and Meta-ComENet, respectively. The same pattern persists at \(20\) shots, where Meta-LegNet still achieves the lowest error.

These results indicate that the combination of self-defined adsorption-environment learning and meta-learning improves both fast adaptation and sample efficiency. The model benefits from additional support examples, but it already retains a clear performance advantage when only a single example is available. This indicates that the model has a stronger extrapolation capability.

\begin{table}[h]
\centering
\scriptsize
\caption{Few-shot adaptation results (MAE, eV) on representative 0D (CH$_2$CO), 2D (OH), and 3D (CHOHCH$_2$OH) adsorption tasks. Lower values indicate better performance.}
\label{tab:few_shot_results_three}
\begin{tabular}{@{}llcccc@{}}
\toprule
\textbf{Model} & \textbf{Dataset} & \textbf{1-Shot} & \textbf{5-Shot} & \textbf{10-Shot} & \textbf{20-Shot} \\
\midrule
\multirow{3}{*}{\textbf{Meta-LegNet}}
& CH$_2$CO & 0.0923 & 0.0787 & 0.0721 & 0.0612 \\
&OH & 0.0828 & 0.0768 & 0.0744 & 0.0699 \\
& CHOHCH$_2$OH & 0.3320 & 0.3118 & 0.2676 & 0.2118 \\
\midrule
\multirow{3}{*}{Meta-CGCNN}
& CH$_2$CO & 0.1369 & 0.1314 & 0.1165 & 0.1084 \\
& OH & 0.2661 & 0.2375 & 0.2109 & 0.1686 \\
& CHOHCH$_2$OH & 0.6874 & 0.6191 & 0.5772 & 0.5182 \\
\midrule
\multirow{3}{*}{Meta-SchNet}
& CH$_2$CO & 0.1289 & 0.1201 & 0.1099 & 0.0971 \\
& OH & 0.2842 & 0.2614 & 0.2476 & 0.1856 \\
& CHOHCH$_2$OH & 0.6308 & 0.5742 & 0.4835 & 0.4571 \\
\midrule
\multirow{3}{*}{Meta-ComENet}
& CH$_2$CO & 0.1075 & 0.0951 & 0.1015 & 0.0894 \\
& OH & 0.1971 & 0.1454 & 0.1145 & 0.0924 \\
& CHOHCH$_2$OH & 0.5142 & 0.4775 & 0.4301 & 0.3972 \\
\bottomrule
\end{tabular}
\end{table}

\subsection{Adsorption-Environment Validation}

Beyond predictive accuracy, an important objective of Meta-LegNet is to identify which atoms constitute the physically relevant adsorption environment. 
Because the model predicts adsorption energy through a local--global decomposition,
\begin{equation}
E_{\mathrm{ads}}^{\mathrm{pred}}=\sum_{i\in V} c_i+\Delta E_{\mathrm{global}},
\end{equation}
each atom can be assigned a learned local contribution $c_i$. 
We convert these contributions into normalized importance scores,
\begin{equation}
\hat{w}_i=\frac{|c_i|}{\sum_{j\in V}|c_j|}, \qquad \sum_{i\in V}\hat{w}_i=1,
\end{equation}
and define the adsorption environment by applying automatic one-dimensional segmentation to $\{\hat{w}_i\}$. 
In practice, Otsu thresholding is used to separate foreground and background atoms:
\begin{equation}
\mathcal{E}=\{\,i\in V : \hat{w}_i \text{ is classified as foreground}\,\}.
\end{equation}

\begin{figure*}[t]
\centering
\includegraphics[width=1\textwidth]{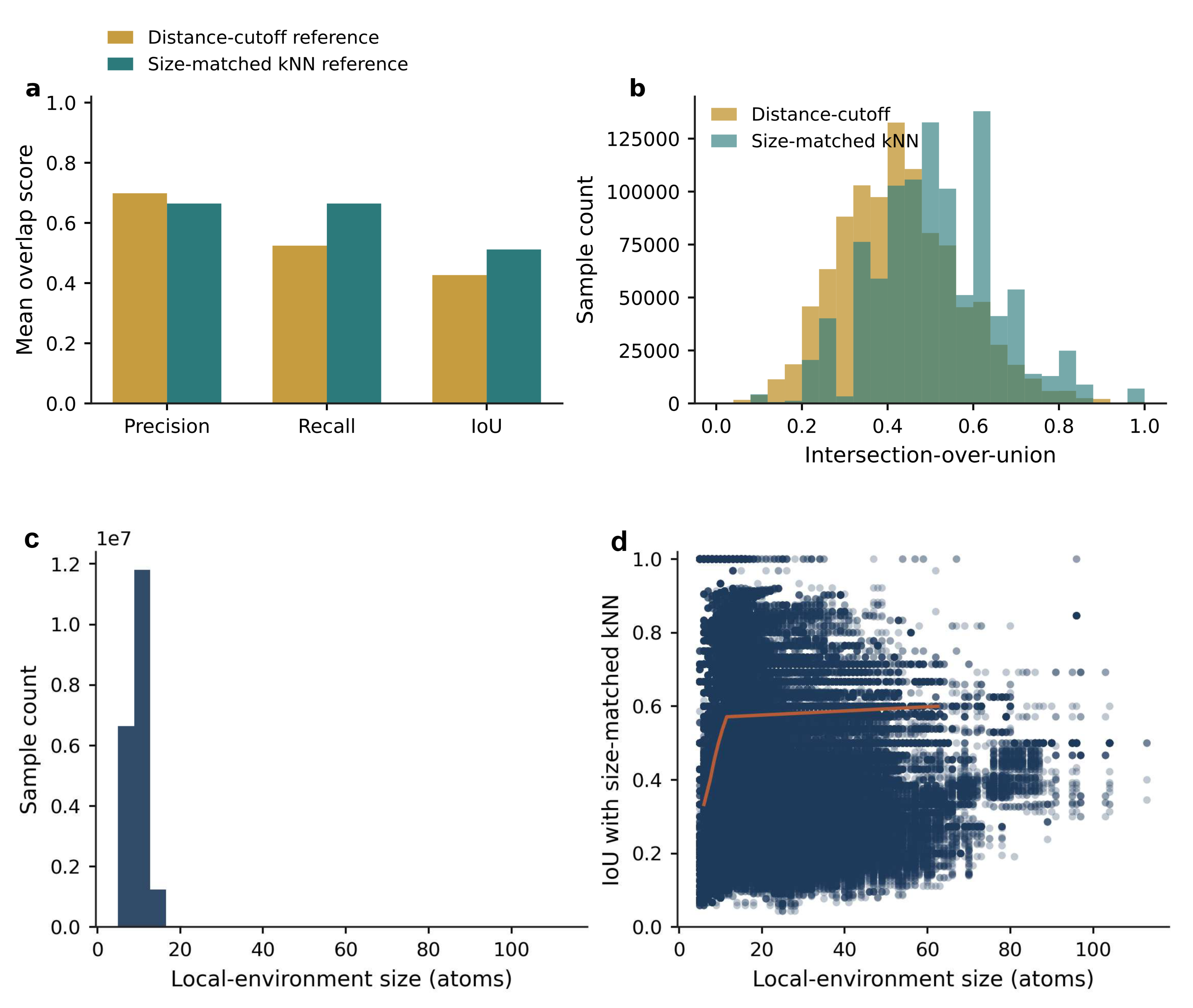}
\caption{Validation of the learned adsorption environments against geometric reference environments. 
(a) Mean precision, recall, and intersection-over-union (IoU) of the learned environments compared with a fixed distance-cutoff reference and a size-matched $k$-nearest-neighbor (kNN) reference. 
(b) Distribution of sample-wise IoU values for the two references. 
(c) Distribution of learned local-environment sizes, showing that the extracted environments remain compact while varying adaptively across adsorption structures. 
(d) IoU of the learned environments with the size-matched kNN reference as a function of local-environment size. 
The learned environments show stronger overall agreement with the size-matched kNN reference than with the fixed distance-cutoff baseline, confirming their local nearest-neighbor-like geometric character. 
However, the overlap is far from complete, indicating that a substantial fraction of adsorption-relevant atoms selected by Meta-LegNet cannot be recovered by simple cutoff or kNN rules.
}
\label{fig:7}
\end{figure*}

We first examine whether the extracted environment is geometrically meaningful by comparing $\mathcal{E}$ with two conventional geometric references: 
(i) a fixed distance-cutoff environment and 
(ii) a size-matched $k$-nearest-neighbor (kNN) environment containing the same number of atoms as the learned environment. 
Figure~\ref{fig:7} summarizes the comparison. 
As shown in Figure~\ref{fig:7}a, the learned environment exhibits stronger overall agreement with the size-matched kNN reference than with the fixed distance-cutoff reference, especially in terms of recall and intersection-over-union (IoU). 
Although the distance-cutoff reference can achieve comparable or slightly higher precision, its lower recall and IoU indicate that it captures only a limited and overly rigid subset of the atoms identified by the model. 
By contrast, the size-matched kNN reference better reflects the compact nearest-neighbor-like structure of the learned environment, suggesting that Meta-LegNet extracts adsorption environments with clear locality and geometric rationality.

The sample-wise IoU distributions in Figure~\ref{fig:7}b further support this conclusion. 
Compared with the distance-cutoff reference, the size-matched kNN reference is shifted toward higher overlap values, indicating that the learned environments are more similar to adaptive local neighborhoods than to fixed-radius shells. 
Importantly, however, the IoU values are not close to unity and remain broadly distributed. 
This means that even when the number of atoms is matched, a simple kNN construction cannot fully reproduce the atoms selected by the model. 
In other words, roughly only part of the learned adsorption environment can be explained by standard geometric proximity, while a substantial non-overlapping fraction cannot be recovered by either cutoff or kNN rules.

We further analyze the size characteristics of the extracted environments. 
Figure~\ref{fig:7}c shows that the learned environments remain compact, with most selected environments containing only a small number of atoms, but their sizes still vary across adsorption systems. 
This indicates that the model does not select diffuse or arbitrary atom sets from the whole structure; instead, it focuses on localized adsorption-relevant regions whose extent adapts to the catalyst--adsorbate configuration. 
Consistently, Figure~\ref{fig:7}d shows that the overlap with the size-matched kNN reference remains substantial over different local-environment sizes, confirming that the learned environments preserve a nearest-neighbor-like geometric character while allowing system-dependent deviations from purely distance-based selection.

These deviations are essential rather than incidental. 
The incomplete overlap with both geometric references indicates that adsorption-relevant atoms cannot be fully identified by a universal radial cutoff or by simply taking the nearest atoms. 
Atoms missed by these hand-crafted rules may still contribute to adsorption through local coordination, surface morphology, adsorbate-induced structural effects, or electronic interactions. 
Therefore, the learned environment should be interpreted as a chemically adaptive local motif: it is geometrically local and physically reasonable, but it is not reducible to a conventional cutoff or kNN definition.

The full distributional results are provided in Figure~S2. 
The IoU and recall histograms consistently show that the learned environments are more closely aligned with the size-matched kNN reference than with the fixed distance-cutoff reference. 
This confirms that the extracted environments have a clear local geometric structure. 
At the same time, the distributions remain broad rather than collapsing near perfect overlap, showing that geometric heuristics can only approximate the learned environment. 
Thus, the learned adsorption environment combines two important properties: it is local and geometrically reasonable, but it also contains adsorption-relevant atoms that are not captured by traditional geometric definitions.

To further examine size effects, Figure~S3 plots IoU and recall as functions of the learned environment size. 
The overlap with the distance-cutoff reference varies strongly with environment size, reflecting the limitation of using a single radial threshold across chemically diverse adsorption systems. 
The size-matched kNN reference provides more stable agreement, but still leaves a substantial non-overlapping fraction across the full size range. 
Taken together, these results show that Meta-LegNet identifies compact local adsorption environments whose spatial organization resembles nearest-neighbor neighborhoods, while their detailed composition is adaptively determined by the adsorption system rather than by a fixed hand-crafted rule.

\begin{figure*}[t]
\centering
\includegraphics[width=0.8\textwidth]{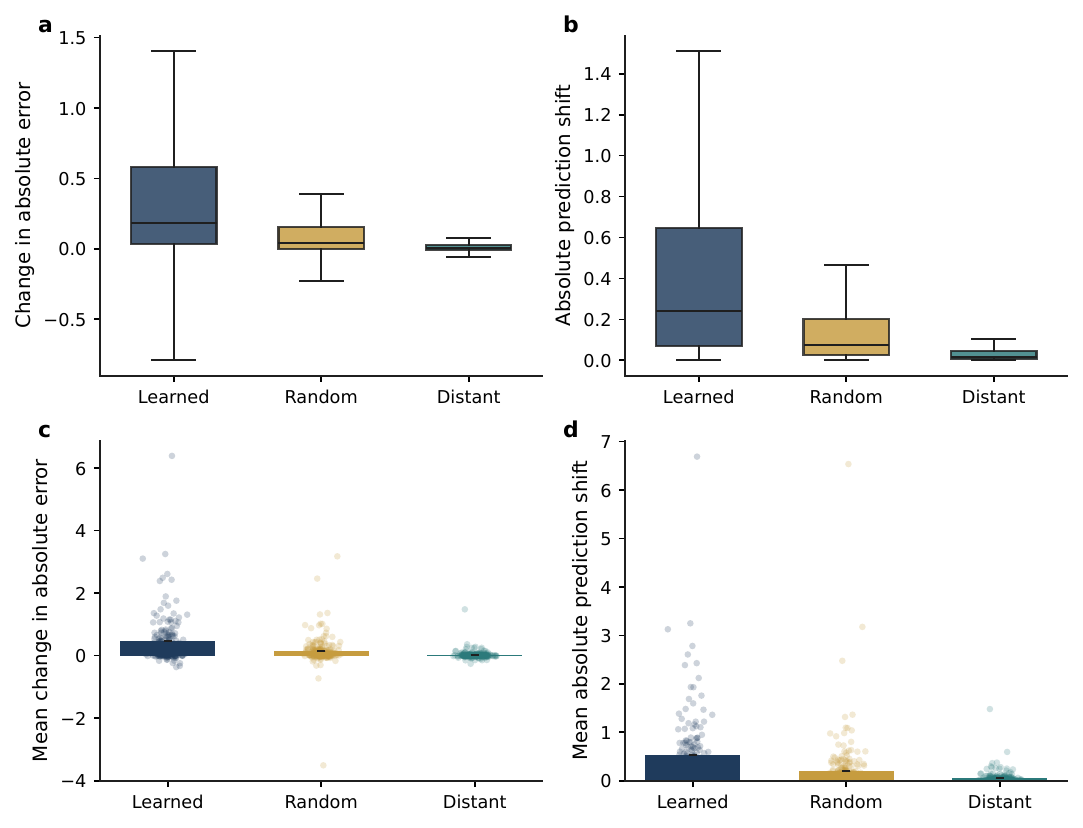}
\caption{Masking-based faithfulness analysis of the learned adsorption environment. 
Removing the learned environment causes a larger prediction shift than removing random or distant control masks of the same size, indicating that the extracted atom set is the most adsorption-relevant subset for the prediction.}
\label{fig:faithfulness}
\end{figure*}

We further evaluate the adsorption relevance of the learned environment through a masking-based faithfulness analysis. 
Starting from the full adsorption structure, we remove the non-adsorbate atoms belonging to a selected environment and recompute the prediction on the perturbed graph. 
A faithful adsorption environment should induce a larger prediction change when removed than a control subset of the same size. 
We therefore compare the learned environment with two controls: a random environment and a distant environment consisting of atoms farthest from the adsorbate.

As shown in Figure~\ref{fig:faithfulness}, removing the learned environment produces the largest change in absolute error and the largest absolute prediction shift. 
Random masks cause smaller perturbations, while distant masks have the weakest effect. 
This result demonstrates that the atoms selected by Meta-LegNet are not merely geometrically close to the adsorbate, but are also the most important atoms for the model's adsorption-energy prediction. 
Therefore, the learned environment is both geometrically meaningful and prediction-faithful.

Taken together, the overlap analysis and masking tests show that Meta-LegNet extracts adsorption environments that are more scientifically effective than traditional cutoff or kNN rules. 
The overlap with the size-matched kNN reference confirms that the learned environments are compact, local, and geometrically reasonable. 
However, the incomplete overlap with both geometric references shows that simple hand-crafted rules miss a substantial fraction of atoms that the model identifies as adsorption-relevant. 
The masking analysis further verifies that these learned atoms make the strongest contribution to the prediction. 
Thus, Meta-LegNet provides an adaptive, data-driven strategy for identifying physically meaningful adsorption environments beyond conventional geometric heuristics.

\begin{figure*}[t]
\centering
\includegraphics[width=0.8\textwidth]{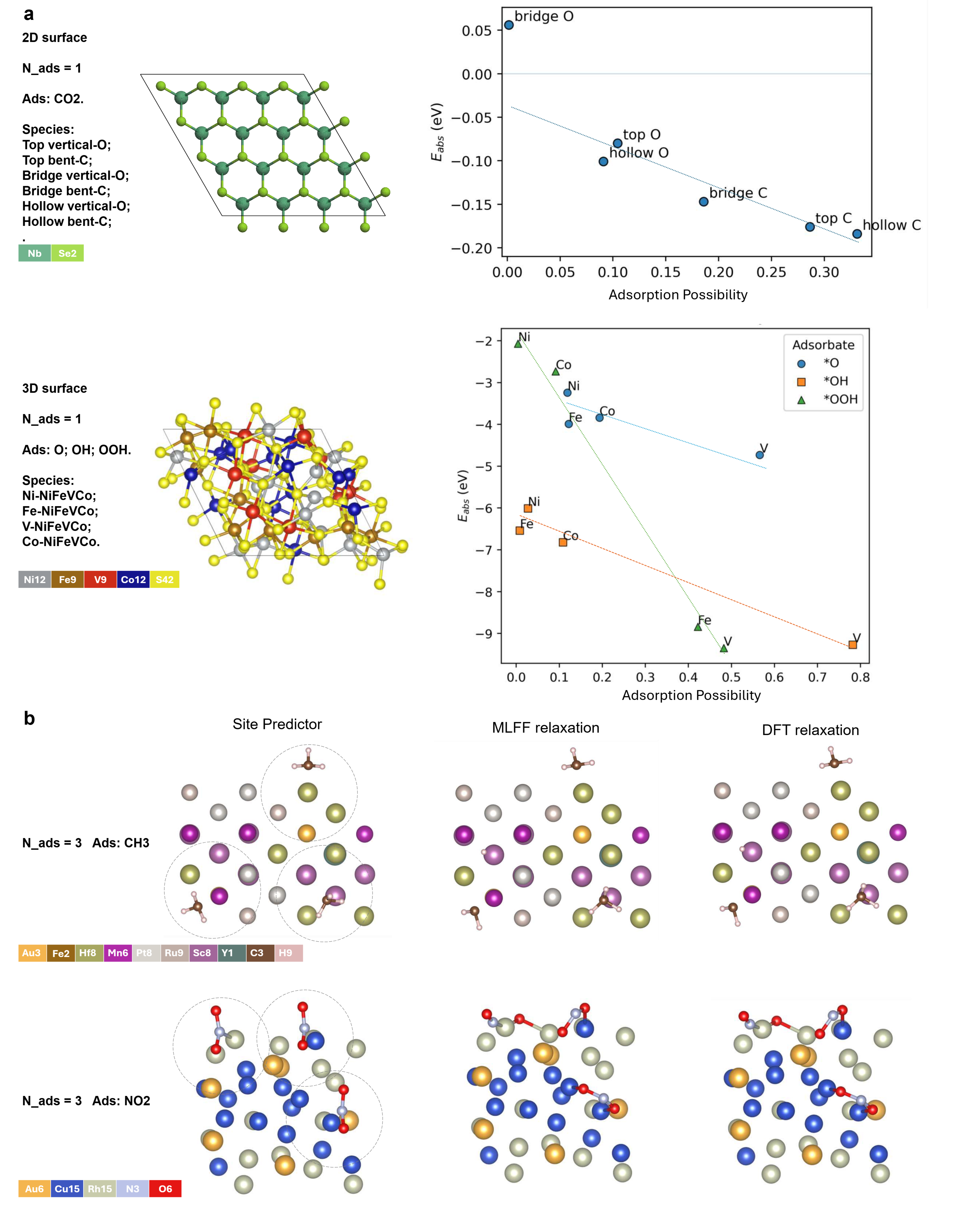}
\caption{SEAM enables direct adsorption-site localization on unseen surfaces.
(a) For representative 2D and 3D surfaces, SEAM predicts feasible adsorption sites and adsorption possibilities across surface motifs and adsorbate species. The predicted probabilities correlate with adsorption energies, indicating that high-probability sites correspond to energetically favorable configurations.
(b) On irregular stepped Pt\(_{192}\)(100)-derived surfaces, created by removing the top five atoms, SEAM proposes adsorption sites and initial poses for multiple adsorbates without exhaustive site enumeration. The MLFF- and DFT-relaxed structures confirm that the predicted sites provide reliable starting configurations for adsorption modeling.}
\label{fig:SEAM}
\end{figure*}

\subsection{Construction of an Adsorption Environment Database and SEAM}

To support scalable use beyond per-sample prediction, we extract all learned adsorption environments across the full dataset and organize them into an adsorption-environment database. Using the segmentation procedure described above, we process approximately 600,000 adsorption scenarios and obtain more than 160,000 unique adsorption environments after structural deduplication.

Rather than storing every extracted environment indiscriminately, we further assess the reliability and reusability of each learned adsorption environment before database construction. Specifically, we retain environments that are supported by three complementary criteria. First, we evaluate stability, i.e., whether the same local environment is consistently identified across repeated model runs or equivalent inference settings. Second, we evaluate overlap with independently constructed geometric references, which measures whether the learned environment remains geometrically consistent with a physically plausible adsorption neighborhood. Third, we evaluate masking faithfulness, which measures whether removing the learned environment leads to a substantially larger degradation in prediction quality than matched random or distant controls. Together, these three criteria allow us to distinguish robust, model-relevant local environments from unstable or incidental ones.

For each retained adsorption environment, the database stores the atomistic structure together with per-atom importance weights, latent representations learned by the model, and a set of derived geometric and chemical descriptors. These descriptors encode not only elemental composition and local coordination geometry, but also the model-derived relevance of individual atoms within the environment. To improve reuse and reduce redundancy, we further perform descriptor-based clustering and structural deduplication, so that adsorption environments with similar geometric and chemical motifs can be grouped into reusable environment families.

Each database entry is then assigned a quality tier according to its combined stability, overlap, and masking scores. In this way, the database does not simply serve as a collection of extracted fragments, but as a filtered library of reliable and reusable adsorption-environment templates. These templates enable downstream similarity analysis, retrieval, interpretation, and transfer across surfaces.

The database enables transfer in two ways. First, it provides reusable prototypes for similarity analysis, few-shot learning, and interpretation. Second, it supports Site Extraction via Adsorption-environment Matching (SEAM), which matches learned environment templates to local motifs on new surfaces using element-aware geometric consistency and rigid registration. As illustrated in Figure~\ref{fig:SEAM}, SEAM returns ranked candidate sites and poses without exhaustive site enumeration. These proposals can be used directly or passed to DFT as high-quality initial configurations for further refinement. In this sense, the database and SEAM extend the value of Meta-LegNet beyond regression accuracy and turn the learned adsorption environments into reusable objects for catalyst discovery workflows.

\section{Conclusion and Discussion}

Meta-LegNet addresses a central bottleneck in computational catalysis: learning
adsorption-relevant local environments that support accurate, transferable, and
interpretable adsorption prediction. By combining SE(3)-equivariant atom-level
message passing, voxel-based multiscale aggregation, assignment-based upsampling,
gated feature fusion, and cross-domain meta-learning, the framework learns
transferable adsorption-environment representations across 0D, 2D, and 3D
adsorption benchmarks. Across the evaluated tasks, Meta-LegNet achieves the lowest MAE
to the baselines and remains effective in few-shot settings, indicating
that the learned representations can transfer across catalyst-adsorbate systems
with limited target-domain data.

The model also provides an interpretable route for adsorption-environment analysis.
The local--global readout generates atom-wise attribution maps, which highlight
regions that contribute strongly to the predicted adsorption energy. These
attribution signals are further processed through adsorption-environment
segmentation, screening, feature extraction, and classification, enabling the
extracted environments to be organized into a reusable adsorption-environment
database. Based on this database, the SEAM procedure proposes plausible adsorption
sites and candidate adsorption configurations on previously unseen surfaces without
requiring exhaustive site enumeration.

Several limitations remain. First, the current study mainly focuses on adsorption
energies and optional force predictions; extending the framework to reaction
barriers, electronic properties, kinetic descriptors, or other catalytic quantities
would broaden its applicability. Second, although Meta-LegNet improves transfer
across multiple material classes, performance under more extreme data scarcity may
still benefit from additional physical priors, such as explicit electronic-structure
features, charge information, or multi-fidelity supervision. Third, the present
datasets do not explicitly include environmental effects such as temperature,
solvent, coverage-dependent dynamics, or surface restructuring, all of which can
affect adsorption behavior under realistic catalytic conditions. 

Future work will therefore focus on three directions: extending the framework to a
broader range of catalytic properties, integrating additional physics and
multi-fidelity information, and incorporating experimental measurements or
operando-relevant conditions where available. Overall, Meta-LegNet provides a
practical route to faster and more interpretable adsorption modeling by shifting the
focus from exhaustive candidate enumeration to transferable adsorption-environment
learning, offering a useful foundation for data-driven catalyst discovery.
\bibliography{arxiv}
\bibliographystyle{unsrt}

\section{Supplementary}

\subsection{S1}

\begin{table}[htbp]
\centering
\caption{Ablation study on the LegNet architecture. MAE is reported in eV, and lower values indicate better performance.}
\label{tab:ablation}
\begin{tabular}{lccc}
\hline
\textbf{Dataset / Task} & \textbf{Vanilla Model} & \textbf{+ Coarsening} & \textbf{Full LegNet} \\
\hline
0D / IS2RE & 0.1140 & 0.1106 & \textbf{0.1052} \\
0D / S2E & 0.1102 & 0.1089 & \textbf{0.1013} \\
2D / IS2RE & 0.0288 & 0.0281 & \textbf{0.0261} \\
3D / IS2RE & 0.1261 & 0.1201 & \textbf{0.0968} \\
OC20-traj / IS2RE & 0.3852 & 0.3621 & \textbf{0.3445} \\
OC20-traj / S2E & 0.2142 & 0.1967 & \textbf{0.1895} \\
\hline
\end{tabular}
\end{table}

The ablation results show that SE(3) graph coarsening alone yields consistent gains across all datasets and that adding multiscale fusion and equivariant upsampling provides further improvement. The effect is particularly clear on IS2RE, where combining coarse structural context with recovered fine local detail leads to the most accurate relaxed-energy prediction.

\subsection{S2}

To complement the aggregate statistics in the main text, we further examine the sample-wise overlap distributions between the learned adsorption environments and two geometry-based reference environments: a fixed distance-cutoff environment and a size-matched $k$-nearest-neighbor (kNN) environment. 
Figure~S2 reports the corresponding histograms of intersection-over-union (IoU) and recall over the validation set.

The learned environments show a clear nearest-neighbor-like character. 
Compared with the fixed distance-cutoff reference, the size-matched kNN reference yields generally higher IoU and recall distributions, indicating that the atoms selected by the model are more consistent with a local nearest-neighbor organization than with a universal radial shell. 
This result supports the geometric validity of the learned adsorption environments: they are not arbitrary atom selections, but compact local motifs that largely follow the spatial locality expected for adsorption-related interactions.

At the same time, the overlap with the size-matched kNN reference is far from perfect. 
Although the kNN reference captures a substantial portion of the learned environments, the distributions remain broad and do not collapse near unity. 
This means that a considerable fraction of atoms identified by the model cannot be recovered by simply selecting the same number of nearest atoms. 
Similarly, the fixed distance-cutoff reference also fails to include many model-selected atoms, especially when the adsorption-relevant environment deviates from a single predefined radial scale. 
Therefore, while the learned environments are local and geometrically reasonable, they cannot be fully described by either a fixed cutoff rule or a simple kNN construction.

This distinction is important for interpreting the role of the learned adsorption environment. 
The atoms missed by the cutoff and kNN references are not necessarily irrelevant; rather, they may encode adsorption-specific chemical contributions associated with surface morphology, coordination pattern, adsorbate identity, or longer-range local interactions. 
Thus, the incomplete agreement with geometric references indicates that simple geometric heuristics can only approximate part of the adsorption environment. 
A data-driven model is required to identify the full set of atoms that contribute to adsorption in a system-dependent manner.

Overall, the distributional analysis shows that the learned adsorption environments possess both locality and chemical adaptivity. 
Their stronger agreement with the size-matched kNN reference confirms that they retain a physically reasonable local nearest-neighbor structure. 
However, the substantial non-overlapping fraction demonstrates that adsorption-relevant atoms cannot be completely obtained from hand-crafted cutoff or kNN rules. 
This further motivates the use of Meta-LegNet to learn adsorption environments directly from data.

\begin{figure*}[t]
\centering
\includegraphics[width=0.8\textwidth]{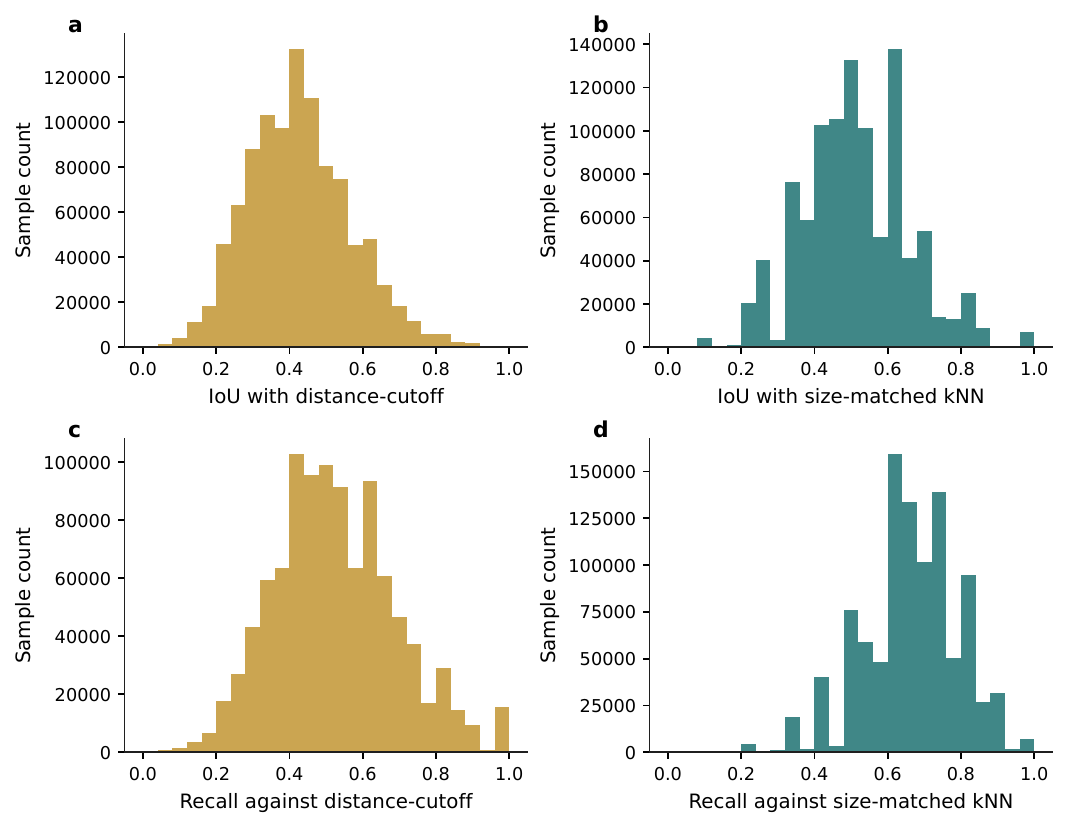}
\caption{Distributional comparison between the learned adsorption environments and geometric reference environments over the full validation set. 
(a) Histogram of IoU with the distance-cutoff reference. 
(b) Histogram of IoU with the size-matched kNN reference. 
(c) Histogram of recall against the distance-cutoff reference. 
(d) Histogram of recall against the size-matched kNN reference. 
The learned environments show stronger agreement with the size-matched kNN reference than with the fixed distance-cutoff reference, indicating that they preserve a local nearest-neighbor-like geometric structure. 
However, the overlap is still far from complete, showing that a substantial fraction of adsorption-relevant atoms selected by the model cannot be recovered by simple cutoff or kNN rules. 
This supports the need for data-driven extraction of adaptive adsorption environments.
}
\label{fig:s2}
\end{figure*}

\subsection{S3}

To further understand how the extracted adsorption environments behave across different spatial scales, we analyze the dependence of overlap statistics on the number of atoms contained in each learned environment. 
Figure~S3 plots IoU and recall against local-environment size for both the distance-cutoff and size-matched $k$-nearest-neighbor (kNN) references.

The size-dependent analysis further confirms the local geometric character of the learned environments. 
Across a broad range of learned environment sizes, the size-matched kNN reference generally maintains higher agreement with the model-selected atoms than the fixed distance-cutoff reference. 
This indicates that, regardless of the absolute number of atoms selected by the model, the learned environments tend to preserve a nearest-neighbor-like local structure. 
In other words, the model does not select atoms randomly from the full structure, but focuses on compact local regions that are geometrically meaningful for adsorption.

However, the scatter plots also show that the learned environments cannot be reduced to either reference definition. 
For the fixed distance-cutoff reference, both IoU and recall vary strongly with environment size, reflecting the limitation of using one universal radial threshold for chemically diverse adsorption systems. 
For the size-matched kNN reference, the agreement is more stable and generally higher, but the overlap remains incomplete across the full size range. 
This means that even when the number of atoms is matched, a simple nearest-neighbor ranking still misses a substantial fraction of atoms selected by the model.

These missing atoms are central to the interpretation of the learned environment. 
They indicate that adsorption-relevant environments are not determined solely by distance or by the number of nearest neighbors. 
Atoms that are not included by the cutoff or kNN references may still contribute significantly to adsorption through system-specific coordination, local geometry, adsorbate-induced interactions, or electronic effects. 
Therefore, the incomplete overlap should not be viewed as a failure of locality, but as evidence that adsorption environments require both local geometric consistency and chemical adaptivity.

These results reinforce the necessity of learning adsorption environments with Meta-LegNet. 
The learned environments are local and geometrically reasonable, as demonstrated by their stronger agreement with size-matched kNN. 
At the same time, they contain adsorption-relevant atoms that cannot be reliably obtained from simple cutoff or kNN heuristics. 
Thus, Meta-LegNet provides an adaptive mechanism for extracting chemically meaningful local environments beyond predefined geometric rules.

\begin{figure*}[t]
\centering
\includegraphics[width=0.8\textwidth]{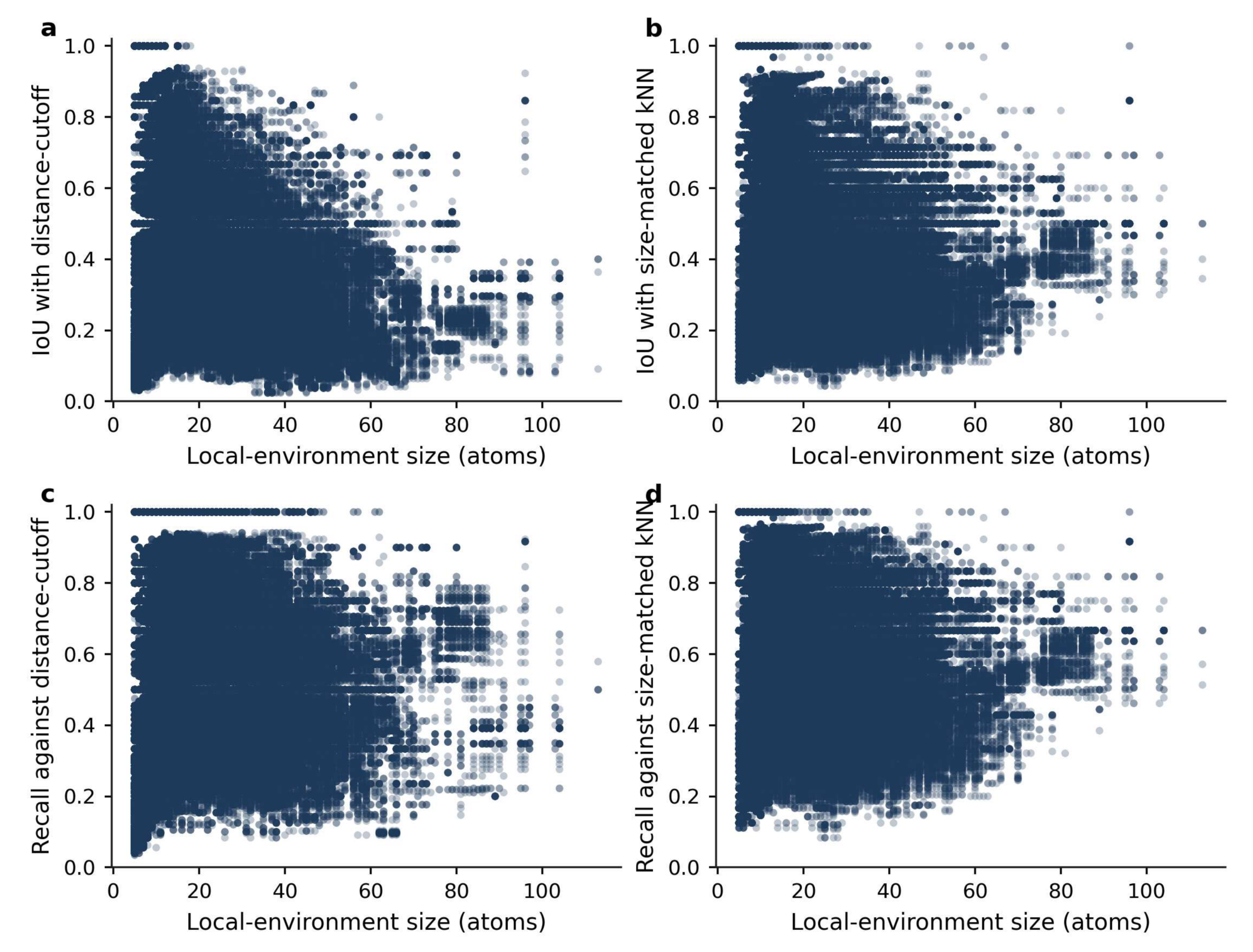}
\caption{Dependence of overlap statistics on the size of the learned adsorption environment. 
(a) IoU with the distance-cutoff reference versus local-environment size. 
(b) IoU with the size-matched kNN reference versus local-environment size. 
(c) Recall against the distance-cutoff reference versus local-environment size. 
(d) Recall against the size-matched kNN reference versus local-environment size. 
The learned environments show stronger and more stable agreement with the size-matched kNN reference than with the fixed distance-cutoff reference across different environment sizes, confirming their local nearest-neighbor-like geometric character. 
Nevertheless, the persistent non-overlapping fraction indicates that many adsorption-relevant atoms selected by the model cannot be captured by simple cutoff or kNN rules, highlighting the need for adaptive environment extraction by Meta-LegNet.
}
\label{fig:s3}
\end{figure*}

\end{document}